\documentclass[sigconf,screen]{acmart}
\usepackage[ruled,linesnumbered,resetcount]{algorithm2e}
\usepackage{tcolorbox}
\usepackage{algpseudocode}
\usepackage[utf8]{inputenc}
\usepackage{amsthm}
\usepackage{amsmath}
\usepackage{caption}
\usepackage{subcaption}

\usepackage{multicol}%
\usepackage{multirow}
\usepackage{xcolor}
\usepackage[normalem]{ulem}

\usepackage{balance}
\usepackage{flushend}

\definecolor{officegreen}{rgb}{0.0, 0.5, 0.0}

\definecolor{stringColor}{rgb}{0.16,0.00,1.00}
\definecolor{fieldColor}{rgb}{0.16,0.00,1.00}
\definecolor{annotationColor}{rgb}{0.39,0.39,0.39}
\definecolor{keywordColor}{rgb}{0.50,0.00,0.33}
\definecolor{commentColor}{rgb}{0.25,0.50,0.37}
\definecolor{javadocColor}{rgb}{0.25,0.37,0.75}
\definecolor{jTagColor}{rgb}{0.50,0.62,0.75}
\definecolor{eTagColor}{rgb}{0.50,0.62,0.75}
\definecolor{lineNumberColor}{rgb}{0.47,0.47,0.47}
\usepackage{listings}
\AtBeginDocument{%
  \providecommand\BibTeX{{%
    \normalfont B\kern-0.5em{\scshape i\kern-0.25em b}\kern-0.8em\TeX}}}

\newcommand{\cgpruner}{\textsc{cgPruner}}
\newcommand{\tool}{\textsc{AutoPruner}}

\setcopyright{acmlicensed}
\acmPrice{15.00}
\acmDOI{10.1145/3540250.3549175}
\acmYear{2022}
\copyrightyear{2022}
\acmSubmissionID{fse22main-p1549-p}
\acmISBN{978-1-4503-9413-0/22/11}
\acmConference[ESEC/FSE '22]{Proceedings of the 30th ACM Joint European Software Engineering Conference and Symposium on the Foundations of Software Engineering}{November 14--18, 2022}{Singapore, Singapore}
\acmBooktitle{Proceedings of the 30th ACM Joint European Software Engineering Conference and Symposium on the Foundations of Software Engineering (ESEC/FSE '22), November 14--18, 2022, Singapore, Singapore}

\begin{document}

\title{\tool{}: Transformer-Based Call Graph Pruning}


\author{Thanh Le-Cong}
\author{Hong Jin Kang}
\author{Truong Giang Nguyen}
\author{Stefanus Agus Haryono}
\author{David Lo}
\affiliation{%
  \institution{Singapore Management University}
  \city{Singapore}
  \country{Singapore}
}





\author{Xuan-Bach D. Le}
\affiliation{%
  \institution{University of Melbourne}
  \city{Melbourne}
  \state{Victoria}
  \country{Australia}
}

\author{Quyet Thang Huynh}
\affiliation{%
  \institution{Hanoi University of Science and Technology}
  \city{Hanoi}
  \country{Vietnam}
}

\renewcommand{\shortauthors}{Thanh C. Le, Hong Jin Kang, Giang T. Nguyen, Stefanus A. Haryono, David Lo, Xuan-Bach D. Le, Thang Q. Huynh}




\begin{abstract}

Constructing a static call graph requires trade-offs between soundness and precision. 
Program analysis techniques for constructing call graphs are unfortunately usually  imprecise. 
To address this problem, researchers have recently proposed \textit{call graph pruning} empowered by machine learning to post-process call graphs constructed by static analysis. A machine learning model is built to capture information from the call graph by extracting structural features for use in a random forest classifier. It then removes edges that are predicted to be false positives. Despite the improvements shown by machine learning models, they are still limited as they do not consider the source code semantics and thus often are not able to effectively distinguish true and false positives.

In this paper, we present a novel call graph pruning technique, \tool{}, for eliminating false positives in call graphs via both statistical semantic and structural analysis. 
Given a call graph constructed by traditional static analysis tools, \tool{} takes a Transformer-based approach to capture the semantic relationships between the caller and callee functions associated with each edge in the call graph.
To do so, \tool{} fine-tunes a model of code that was pre-trained on a large corpus to represent source code based on descriptions of its semantics.
Next, the model is used to extract semantic features from the functions related to each edge in the call graph. 
\tool{} uses these semantic features together with the structural features extracted from the call graph to classify each edge via a feed-forward neural network.
Our empirical evaluation on a benchmark dataset of  real-world programs shows that \tool{} outperforms the state-of-the-art baselines, improving on F-measure by up to 13\% in identifying false-positive edges in a static call graph. 
Moreover, \tool{} achieves improvements on two client analyses, including halving the false alarm rate on null pointer analysis and over 10\% improvements on monomorphic call-site detection.
Additionally, our ablation study and qualitative analysis show that the semantic features extracted by \tool{} capture a remarkable amount of information for distinguishing between true and false positives.

\end{abstract}





\maketitle

\section{Introduction}\label{sec:intro}
Call graphs construction is crucial for program analyses. 
Call graphs capture invocations between functions of programs \cite{callahan1990constructing, ryder1979constructing}. 
An ideal call graph  is (1) sound, i.e., it does not miss any true invocations, and (2) precise, i.e., it does not produce any false positives. 
However, even for small programs, constructing a sound and precise call graph is difficult~\cite{ali2012application}. 
A call graph analysis should make reasonable trade-offs between soundness and precision. 
Unfortunately, recent work~\cite{utture2022striking} has found that widely used tools such as WALA \cite{fink2012wala}  or Petablox \cite{mangal2015user} construct imprecise call graphs; up to 76\% of edges in  call graphs constructed by WALA are false positives. 


To address these issues, researchers \cite{bravenboer2009strictly, mangal2015user, tan2016making} have proposed to improve pointer analysis, which is the core of many call graph constructions algorithms, by improving context-sensitivity or flow-sensitivity of the analysis. 
Unfortunately, a perfect pointer analysis is generally not possible \cite{rice1953classes}. Specifically, pointer analyses usually face an expensive trade-off between scalability and precision  \cite{li2018scalability}. For example, a context-sensitive analysis by WALA only reduces 8.6\% false positives rate over a context-insensitive analysis while incurring a large performance overhead~\cite{utture2022striking}. 

A recent approach, which we refer to as \cgpruner{} \cite{utture2022striking}, achieved a breakthrough in improving the quality of call graphs.
Instead of directly improving pointer analysis, \cgpruner{} performs call graph \textit{pruning}  as a post-processing technique on a  call graph constructed through static analysis. 
Using machine learning techniques, the call graph pruner removes false positives in a call graph.
Specifically, \cgpruner{} first extracts a set of \emph{structural} features from the call graph, e.g., the number of outgoing edges from the call-site and the in-degree of the callee. It then leverages a learning model, i.e., Random Forest, to predict if a caller-callee edge is a false positive (i.e., the caller does not invoke the callee in reality). 
The call graph is updated by removing edges that are predicted to be false positives. Since the cost of generating predictions is low, a machine learning-based approach does not incur a significant performance overhead. Their experiments show that \cgpruner{} successfully improves over traditional call graph analysis by producing call graphs that eliminate a large number of false positives. However, \cgpruner{} is still limited as it does not consider source code semantics and thus is not able to distinguish true and false postives effectively.

\begin{figure*}[!htb]
\centering
    \includegraphics[width=0.7\textwidth]{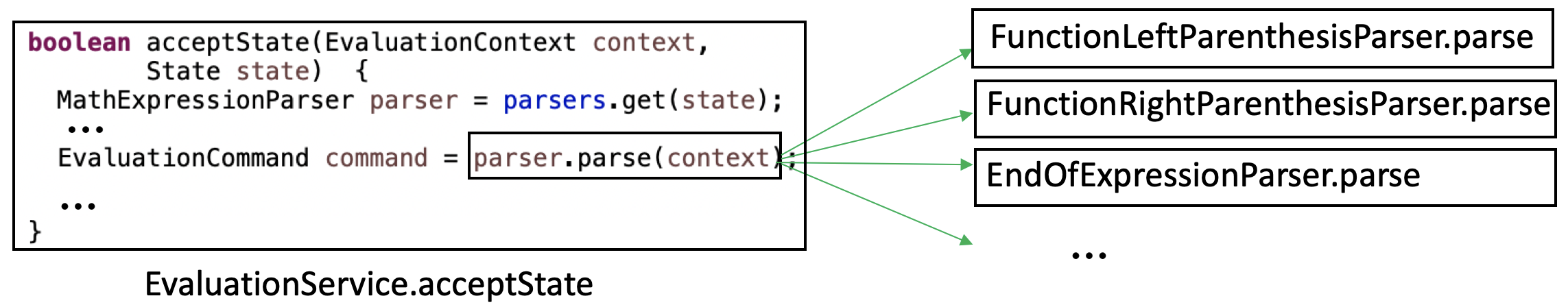}
\caption{The \texttt{parser.parse} call depends on the parameter \texttt{state}. \texttt{parsers.get} returns an object of one of multiple  classes implementing \texttt{MathExpressionParser}. Throughout the course of the program execution, \texttt{acceptState} can be invoked  with all possible values of \texttt{state}. Therefore,
despite the large number of outgoing edges from the same call site, all edges to the \texttt{parse} nodes from \texttt{acceptState} in the statically computed call graph are true edges. However, the large number of outgoing edges is a feature used to prune false positives, which leads to the incorrect pruning of the edges related to the \texttt{parse} calls.} 
  \label{fig:polymorphic_callsite}
\end{figure*}

In this paper, we propose a novel technique, \tool{}, that incorporates both structural and statistical semantic information to prune false positives in call graphs effectively. 
\tool{} combines structural features extracted from a call graph with semantic features extracted from the source code of the caller and callee functions. Similar to \cgpruner{}, \tool{} uses handcrafted structural features. However, different from \cgpruner{}, \tool{} automatically extracts semantic features via deep learning. \tool{} thus faces a unique challenge on how to use deep learning to automatically learn from a limited dataset.  
To address this, we leverage recently proposed transformer models of code, i.e., CodeBERT \cite{feng2020codebert}, that has been pre-trained on a corpus containing millions of code functions. As our task differs from CodeBERT's pre-training task, \tool{} first fine-tunes CodeBERT such that it captures the statistical relationships between caller and callee functions, learning to distinguish between true and false positive edges based on their source code. Next, \tool{} leverages the fine-tuned model to extract \textit{semantic features} of each edge, based on the source code of the caller and callee functions. Each edge then has
an embedding that represents the semantic relationship between the caller and the callee.
For each embedding, \tool{} combines it with the handcrafted structural features to obtain a representation for each edge. Based on this representation, \tool{} employs a neural classifier to classify each edge in a call graph as a true or false positive. 

We evaluate \tool{} on call graphs produced by three well-known tools, i.e., WALA \cite{fink2012wala}, Doop \cite{bravenboer2009strictly}, and Petablox \cite{mangal2015user}, for real-world programs taken from the NJR-1 benchmark \cite{palsberg2018njr} in the same setting of \cgpruner{} \cite{utture2022striking}. 
We compare the call graph generated by \tool{} against multiple baselines, 
including the original call graphs produced by WALA, Doop and Petablox, the call graphs pruned by state-of-the-art approach \cgpruner{}, 
as well as a graph neural network model, 
that applies deep learning to the call graphs.
The latter two baselines consider only structural information. 
Our experiments show that 
\tool{} improves over the state-of-the-art approach by up to 13\% in F-measure. 
Our experiments demonstrate that the use of the semantic features extracted by the transformer-based model enables \tool{} to outperform approaches that consider only structural information. 

We investigate the effect of pruned call graphs produced by \tool{} on client analyses,
which take the call graphs as input to perform other analyses on the programs.
We investigate two client analyses: null pointer analysis and monomorphic call-site detection. 
On null pointer analysis, call graphs pruned by our approach \tool{} lead to significantly reduced false alarm rate of only 12\% while the call graphs pruned by \cgpruner{}~\cite{utture2022striking} and the call graphs constructed by WALA~\cite{fink2012wala} have false alarm rates in null pointer analysis of 23\% and 73\%. 
On monomorphic call-site detection, \tool{} improves over \cgpruner{} by over 8\% in terms of F-measure. 

To better understand \tool{}, we also perform qualitative analysis on its performance.
We leverage t-SNE \cite{van2008visualizing} to visualize the embedding of the call graph edges in a two-dimensional space. 
We find that the semantic features can separate the true and false-positive edges, demonstrating that \tool{} captures a remarkable amount of information from the source code associated with each call graph edge.

In summary, we make the following contributions:

\begin{itemize}
    \item We introduce \tool{}, a novel call graph pruner that uses both code and structural feature to identify false-positive edges in a call graph.
    
    \item We empirically demonstrate that pruned call graph produced by our approach can help analysis tool significantly improve the false alarm rate and F-measure. Notably, in the analysis client of null pointer analysis, \tool{} leads to over 150 more reported warnings while decreasing false alarm rate from 73\% to 12\%. 
    
    \item We perform an ablation study and qualitative analysis to better understand our approach. Our analysis validates the use of our proposed approach for call graph pruning.
\end{itemize}

The paper is structured as follows: Section \ref{sec:background} introduces the background of our work. 
Section \ref{sec:method} describes our proposed approach.
Section \ref{sec:eval} presents our experimental setup and results. 
Section \ref{sec:threats} discusses our qualitative analysis and threats to validity.
Section \ref{sec:related} covers related work. Finally, Section \ref{sec:conclusion} concludes and describes future directions.

\section{Background} \label{sec:background}
In this section, we discuss a motivating example. 
Next, we present the formal formulation of the call graph pruning problem and introduce Transformer-based models of code and CodeBERT \cite{feng2020codebert}. 

\subsection{Motivating Example} \label{sec:motivating}

In Figure \ref{fig:polymorphic_callsite}, we present a motivating example to motivate our approach and demonstrate the limitations of \cgpruner{} that uses only structural features to prune call graphs.
The source code of the \texttt{acceptState} function contains a call to the \texttt{parse} function.
In the original unpruned call graph, the \texttt{acceptState} node is connected to multiple \texttt{parse} nodes of classes that implement the \texttt{MathExpressionParser} interface, e.g. the interface is implemented by
\texttt{EndOfExpressionParser}, \texttt{FunctionLeftParenthesisParser}, and other classes that also override \texttt{accept}.
At runtime, the \\ 
\texttt{acceptState} function is invoked multiple times with different values of \texttt{state}, resulting in calls between \texttt{acceptState} and the multiple \texttt{parse} nodes.
These edges in the call graph are true edges as they represent calls that occur at runtime.

Due to a large number of outgoing edges from the same call site, the local structure of each \texttt{acceptState} to \texttt{parse} edge in the call graph resembles edges that are false positives.
A structural-only approach such as \cgpruner{}, therefore, incorrectly prunes all the edges between the \texttt{acceptState} node and the \texttt{accept} node in the call graph.
This highlights the limitation of considering only the structural features of the call graph and motivates the need for guiding call graph pruning with the semantics of the source code.
From analyzing the source code, we can correctly identify that a large number of outgoing edges are possible, as the specific \texttt{parse} call depends on the parameters of the function. 
Indeed, \tool{} correctly leaves the edges unpruned due to its use of semantic features extracted from the source code by CodeBERT.

\subsection{Call Graph Pruning}
\subsubsection{Problem Formulation}
In this work, we formulate the call graph pruning problem as below.

\vspace{1mm}

\noindent \textbf{Input:} A static call graph $\mathcal{G} = (V, E)$ is a directed graph constructed by a static analysis tool, where $V$ is the set of program's functions identified by a function signature and $E$ is the set of edges, i.e., function calls, in the call graph. 
Each edge in $E$ is a tuple of (caller, callee, offset), where caller is the calling function, callee is the called function, and offset is the call-site in caller.

\vspace{1mm}

\noindent \textbf{Output:} A pruned call graph $\mathcal{G'} = (V', E')$, where $V' = V$ and $E' \subseteq E$ 


\vspace{1mm}



\noindent To address this problem, we aim to train a binary classifier $\mathcal{C}$, which can classify each edge in a call
graph $\mathcal{G}$ as \textbf{true positive}, i.e., the edge represents a true call, or \textbf{false positive}, i.e., the edge does not represents a true call.
Using the classifier's output, we prune the call graph following Algorithm \ref{alg:pruner}.
We use the classifier $\mathcal{C}$ to classify each edge in a call graph. 
Edges classified as false positives are pruned, while edges classified as true positives are retained.

\begin{algorithm}[h]
	\KwIn{Call Graph $\mathcal{G} = (V,E)$, Classifier $\mathcal{C}$
	}
	\KwOut{Pruned Call Graph $\mathcal{G} = (V', E')$}
	\BlankLine
	$\mathcal{G'} \gets \mathcal{G}$ \\
	\ForEach{$e$ in $\mathcal{G}$}{
	    $p \gets \mathcal{C}(e)$ \Comment{prediction of binary classifier} \\
	    \If{$p == $ False-positive}{
	          $E' = E' \setminus \{e\}$ \Comment{remove edge from call graph} \\
	    }
	}
	\Return $\mathcal{G'}$
	\caption{Call-graph Pruner}
	\label{alg:pruner}
\end{algorithm}

\subsubsection{State-of-the-art} 
To prune call graphs, Utture et. al.~\cite{utture2022striking} recently proposed \cgpruner{}, which uses a machine learning model, a random forest classifier, based on 11 structural features extracted from a call graph for pruning edges. 
Their experiments demonstrate that \cgpruner{} could successfully boost the precision of call graphs from 24\% to 66\% and reduce the false positive rate in client tool from 73\% to 23\%. 
The approach, however, also substantially reduces the call graph's recall. 

\cgpruner{} relies only on structural features, which can not distinguish false-positive and true-positive edges that share the same characteristics of structure (as mentioned in Section \ref{sec:motivating}). 
Similar to \cgpruner{}, our approach, i.e. \tool{}, employs a machine learning model to prune call graphs. 
However, unlike \cgpruner{}, we use the semantic information from the source code of both the caller and callee function of an edge in the call graph. 
The information from the source code enables our approach to distinguish true-positive edges from false positive edges (later demonstrated in Section \ref{sec:qualitative}).

\subsection{Transformer Models of Code and CodeBERT}

Transformer models~\cite{vaswani2017attention} are deep learning models based on an encoder-decoder architecture. 
Transformer models employ the attention mechanism and have achieved remarkable performance in field of Natural Language Processing (NLP)~\cite{vaswani2017attention, devlin2018bert, liu2019roberta, yang2019xlnet, radford2019language, brown2020language}. 
NLP models have also been employed for source code-related tasks, as source code has been found to exhibit characteristics, such as repetitiveness and regularity, similar to natural language~\cite{hindle2016naturalness}.

Recently, CodeBERT \cite{feng2020codebert} was proposed as a Transformer model for source code. 
A CodeBERT model is pre-trained on a large corpus, containing over 6 million functions and 2 million pairs of comment-function.
As input, CodeBERT can be given a pair of data (e.g., source code and a code comment that describes the semantics of the source code)
to learn statistical relationships between the pair of data.
CodeBERT is pre-trained by two tasks, Masked Language Modeling (MLM) and Replaced Token Detection (RTD). 
In the MLM task, given an input sequence with a single token, CodeBERT has to predict the value of the masked token. 
In the RTD task, given an input sequence where some tokens are replaced with alternative tokens, CodeBERT detects the replaced tokens.
Previous studies have demonstrated the effectiveness of CodeBERT in multiple tasks, including the capability for CodeBERT to be fine-tuned for tasks that it was not initially trained for~\cite{feng2020codebert, zhou2021finding}. Prior studies have been built on top of CodeBERT for automating various tasks that require an understanding of program semantics, e.g., type inference \cite{kazerounian2021simtyper, hellendoorn2018deep, peng2022static}, program repair \cite{mashhadi2021applying}, etc. Motivated by these success cases, our proposed solution (AutoPruner) built upon CodeBERT for another task, namely call-graph pruning. Our solution tries to capture the semantic relevance of code in the caller function to that of the callee function to differentiate between true and false edges in a call graph. Different from \cgpruner{}, AutoPruner analyzes the source code of the functions while \cgpruner{} ignores them. 
Moreover, AutoPruner considers both caller and callee functions while other pure program analysis methods analyze only the caller function.

\section{Methodology}\label{sec:method}
\begin{figure*}
    \centering
    \includegraphics[width=\textwidth]{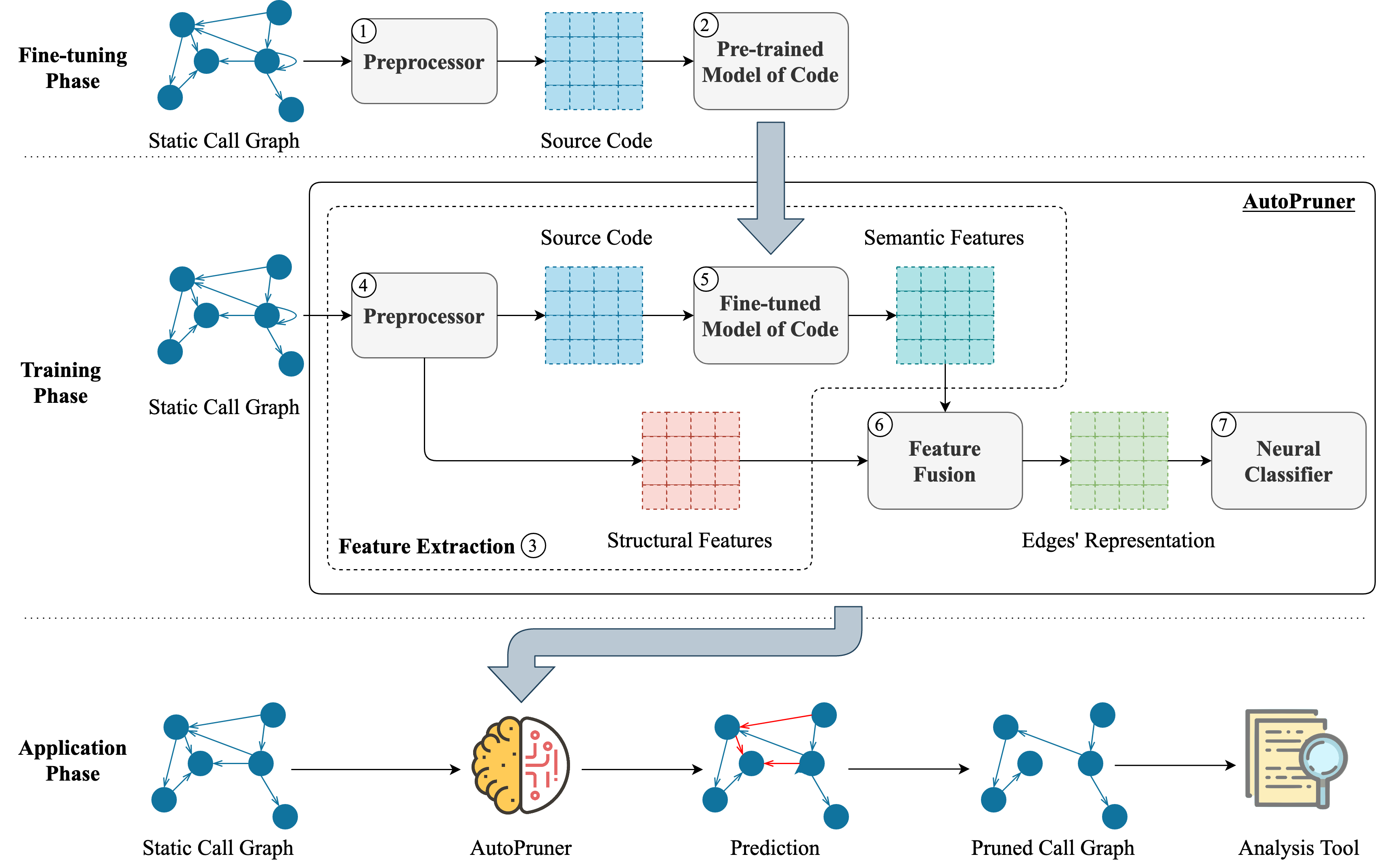}
    \caption{The overview of \tool{}}
    \label{fig:overview}
\end{figure*} 

Figure \ref{fig:overview} illustrates the overall framework of \tool{}. 
Before \tool{} can be applied, it has to be fine-tuned and trained. 
First, in the Fine-tuning phase, we fine-tune a pre-trained CodeBERT model of code to enable it to extract the \textit{semantic features} of the edges in a call graph. 
Next, in the Training phase, we use the fine-tuned model to extract \textit{semantic features} for the edges in the call graph. 
The semantic features are then combined with \textit{structural features} extracted from the call graph to construct the representation of each edge. 
Then, we train a feed-forward neural network classifier to identify false positive edges.
Afterward, in the Application phase, \tool{} can be used as a call graph pruner for post-processing call graphs to be used in other program analyses.

\subsection{Fine-tuning} \label{sec:finetune}
In this phase, we fine-tune CodeBERT.
Before CodeBERT can be used for a  task different from its pretraining task, it has to be fine-tuned to adapt its weights for the new task.
CodeBERT takes a pair of data as input, and in \tool{}, we pass the source code of the caller and callee functions associated with each edge as input.
Specifically, \tool{} uses a preprocessor 
that extracts the source code of the caller and callee functions, constructing a sequence of input tokens that matches the input format expected by CodeBERT. 
Then, the sequences are input into CodeBERT for fine-tuning. 
Below, we explain each component of the fine-tuning phase in detail.

\subsubsection{Pre-processing.} 
The pre-processing step (\textcircled{1} in Figure \ref{fig:overview}) produces the input sequences to CodeBERT, enabling it to learn a representation of the semantic relationship between the source code of the caller and callee functions. 
Initially, from the call graph constructed by a static analysis tool (e.g., WALA or Doop), 
an edge in the call graph is characterized by a pair of function signatures identifying the caller and callee functions associated with it. 
We use \textit{java-parser}\footnote{\url{https://javaparser.org/}} to extract the source code of both the caller and callee functions. Particularly, we parse the source code to obtain the method descriptors for every methods and matches them against the output of existing CG generators, which identifies methods using their descriptors. This allows us to link the source code to the methods in the call graph. Then, we use CodeBERT tokenizer\footnote{\url{https://huggingface.co/microsoft/CodeBERT-base/tree/main}} to tokenize the source code. 
Finally, following the input format of CodeBERT, we construct an input sequence that encodes the source code of the caller and callee functions in the form of:
\begin{equation}
    [CLS] \langle \text{caller's source code} \rangle [SEP] \langle \text{callee's source code} \rangle [EOS]
\end{equation}
where [CLS], [SEP], and [EOS] are tokens separating the pair of data, as required by CodeBERT.

\subsubsection{CodeBERT Fine-tuning.} 
As the pre-training tasks of CodeBERT, (i.e., the Masked Language Modeling and Replaced Token Detection tasks) differs from our task (i.e., identifying false positive edges),
we perform a fine-tuning step to adapt the pre-trained CodeBERT model for our task, following the common practice in transfer learning \cite{torrey2010transfer, zhuang2020comprehensive} and other applications of CodeBERT~\cite{zhou2021finding, mashhadi2021applying, applis2021assessing, zhou2021assessing}.
This step aims to transfer the knowledge based on the pre-training task associated with an extremely large amount of data onto our task where collecting data is expensive (as obtaining the ground truth labels in our task  requires careful human analysis and the execution of test cases).

\vspace{0.2cm}
We fine-tune the CodeBERT model directly on the training dataset of our task of identifying false positive edges in the call graph.
Specifically, we feed input sequences in the input format of CodeBERT, obtained from pre-processing step into the CodeBERT model.
Next, the model extract features from the input sequences. 
Then, we pass the extracted features into a fully connected layer to classify each edge in the call graph into true and false positive edges (\textcircled{2} in Figure \ref{fig:overview}). 

During the fine-tuning phase, the parameters of the CodeBERT model are updated by Adam optimizer \cite{adam} to minimize the Cross-Entropy Loss. After this fine-tuning phase, we freeze all parameters of the CodeBERT model. 
In the subsequent phases, \tool{} uses the fine-tuned CodeBERT model to extract semantic features from the source code of the caller and callee functions associated with each edge.

\subsection{Training}
\label{sec:training}
In the training phase, 
our objective is to train the binary classifier that predicts if a given edge is a true positive or false positive (the Classifier C in Algorithm \ref{alg:pruner}).
To this end, we construct the representation of an edge in a call graph by extracting  and combining features of both types: semantic features (extracted from the source code by fine-tuned language model) and structural features (extracted from the call graph). 
Then, we train a neural classifier to predict whether an edge is true or false positive. 
Below, we explain each component of the pipeline.

\begin{table}[]
    \centering
    \caption{Types of structural features}
    \label{tab:struc_feats}
    \begin{tabular}{|l|l|}
        \hline
        \textbf{Feature} & \textbf{Description}  \\
        \hline
        \textit{src-node-in-deg} &  number of edges ending in caller \\
        \textit{src-node-out-deg} &  number of edges out of caller \\
        \textit{dest-node-in-deg} &  number of edges ending in callee \\
        \textit{dest-node-out-deg} &  number of edges out of callee \\
        \textit{depth} &  length of shortest path from main to caller \\
        \textit{repeated-edges} &   number of edges from caller to callee \\
        \textit{L-fanout} &  number of edges from the same call-site \\
        \textit{node-count} &  number of nodes in call graph \\
        \textit{edge-count} &  number of edges in call graph \\
        \textit{avg-degree} &  average src-node-out-deg in call graph \\
        \textit{avg-L-fanout} &   average L-fanout value in call graph \\
        \hline

    \end{tabular}
    
\end{table}
\subsubsection{Feature Extraction}
We extract into a feature set (\textcircled{3} in Figure \ref{fig:overview}) the two types of features as follows:
\begin{itemize}
    \item \textbf{Semantic features.} The semantic features are extracted from the source code of caller and callee functions using our fine-tuned CodeBERT model.
    To capture this information, we first apply the same pre-processing step as described in the Fine-tuning step (\textcircled{4} in Figure \ref{fig:overview}). Next, the fine-tuned CodeBERT model extracts a high-dimensional vector that encodes the statistical relationship between the caller and callee function (\textcircled{5} in Figure \ref{fig:overview}). As a result, we obtain semantic features of an edge in call graph as follows: 
    \begin{equation}
        f_{sem} = \left\langle v^{sem}_{1}, v^{sem}_{2}, ..., v^{sem}_{k_{c}} \right\rangle
    \end{equation}
    where $k_{c} = 768$ is the embedding dimension of CodeBERT. 
    
    \vspace{0mm}
    
    \item \textbf{Structural features.} The structural feature captures graphical information related to each edge. The features include metrics about the neighborhood of the edge (local information) or the entire call graph  (global information). We use the same features proposed by Utture et al.~\cite{utture2022striking}. Detailed information of the features is presented in Table \ref{tab:struc_feats}. We represent the structural features of an edge in a call graph as follows: 
    \begin{equation}
        f_{struct} = \left\langle v^{struct}_{1}, v^{struct}_{2}, ..., v^{struct}_{k_{s}} \right\rangle
    \end{equation}
    where $k_{s} = 22$ is the number of structural features. Based on the work by Utture et al.~\cite{utture2022striking}, there are two features of each type listed in Table \ref{tab:struc_feats}, one for transitive calls and one for direct calls, so we have 22 structural features.
    
\end{itemize}

\subsubsection{Feature Fusion}
In this step, we combine semantic features and structural features of each edge in the call graph into a final representation. 
We first use one fully connected layer for each feature (\textcircled{6} in Figure \ref{fig:overview}).
Then, the output of these layers are concatenated to produce the final representation. 
More formally,
\begin{equation}
    f'_{sem} = FCN_{k_{c} \times h}(f_{sem})
\end{equation}
\begin{equation}
    f'_{struct} = FCN_{k_{s} \times h}(f_{struct})
\end{equation}
\begin{equation}
        f = \left\langle f'_{sem}, f'_{struct}  \right \rangle
\end{equation}
where $FCN_{m \times n}$ denotes a fully connected layer that takes a $m$-dimensional input and outputs a $n$-dimensional vector. We set $h$, which is the size of the hidden feature vector, as 32. $k_{c}$ and $k_{s}$ is the size of semantic and structural feature vector, respectively.



\subsubsection{Neural Classifier}
Given the representation of an edge obtained from feature extraction, we obtain a score that approximates the probability that an edge is a true positive.

The score is computed through a feed-forward neural network (\textcircled{7} in Figure \ref{fig:overview}) consisting of one hidden layer and one output layer. More formally,
\begin{equation}
    f' = FCN_{2h \times 2}(f)
\end{equation}
\begin{equation}
    prob = OutLayer(f')
\end{equation}
where, $FCN_{m \times n}$ denotes a fully connected layer inputs a $m$-dimensional vector and outputs a $n$-dimensional  one, $OutLayer$ is a softmax function~\cite{goodfellow2016deep}. $f$ and $f'$  are the edge representation and output features, respectively. $prob = \{ prob_{FP}, prob_{TP} \}$ is the output probabilities, where $prob_{FP}$ and $prob_{TP}$ is the probability that an edge is false positive and true positive, respectively. 
An edge with $prob_{TP}$ larger than $prob_{FP}$ is considered as a true positive. 
Otherwise, the edge are considered as a false positive.

During the training phase, the parameters of \tool{} except CodeBERT's parameters are updated by Adam optimizer~\cite{adam} to minimize the cross-entropy loss~\cite{goodfellow2016deep}. 

\subsection{Application}
After the Training Phase, 
\tool{} can now be deployed for use as a call graph pruner.
Given a call graph generated by a static analysis tool, \tool{} preprocesses the call graph and extracts semantic and structural features.
Based on these features, the neural classifier produces predictions for each edge in the call graph.
Using the predictions of the neural classifier, \tool{} removes the edges predicted to be false positives, and 
outputs an improved call graph with fewer false positives.

\tool{} can be integrated into other static analyses (i.e., client analyses) that takes a call graph as input. 
Since call graphs pruned by \tool{} have fewer false positives compared to the original ones, the performance of the client analyses should improve given the more precise call graphs.

\section{Empirical Evaluation}\label{sec:eval}
\subsection{Research Question}
Our evaluation aims to answer the following research questions: 

\vspace{0mm}

\noindent \textbf{RQ1:} \textit{Is \tool{} effective in pruning false positives from static call graphs?} 
This research question concerns the ability of \tool{} in identifying false positive edges in a static call graph. 
To evaluate our approach, we evaluate \tool{} on a dataset of 141 real-world programs in NJR-1 dataset \cite{palsberg2018njr} in terms of Precision, Recall, and F-measure. 
We compare our approach to multiple baselines, 
including the state-of-the-art technique, \cgpruner{}~\cite{utture2022striking}, as well as 
a graph neural network, 
and the original call graphs produced by static analysis tools. 

\vspace{0mm}

\noindent \textbf{RQ2:} \textit{Can \tool{} boost the performance of client analyses?}
This research question investigates the impact of pruned call graphs produced by \tool{} on client analysis.
To answer this question, we use pruned call graph as input for client analyses and investigate its performance compared to original call graph and \cgpruner{}~\cite{utture2022striking} on two client analyses used by \cgpruner{}: null-pointer analysis and monomorphic call-site detection.

\vspace{0mm}

\noindent \textbf{RQ3:} \textit{Which components of \tool{} contributes to its performance?}
\tool{} uses multiple types of features, including the semantic features extracted from both the caller and callee functions, and the structural features extracted from the call graph.
We investigate the contribution of each feature in an ablation study by dropping each type of feature and observing the change in \tool{}'s performance.

\vspace{0mm}

\subsection{Experimental Setup}
\subsubsection{Dataset} To evaluate effectiveness of our approach, we use a dataset of 141 programs, initially constructed by Utture et al.~\cite{utture2022striking}, from the NJR-1 benchmark suite~\cite{palsberg2018njr}.
We follow the same experimental setting as  prior work~\cite{utture2022striking}. 

This dataset of programs was curated by Utture et al.~\cite{utture2022striking} based on the criteria that each program has over 1,000 functions, has over 2,000 call graph edges, and has over 100 functions that are invoked at runtime, 
and has a high overall code coverage (68\%). 
In total, the dataset comprises over 860,000 call graph edges. The ground-truth label of each edge was obtained based on instrumentation and dynamic analysis~\cite{utture2022striking}. 
We use 100 programs as our training set and the remaining programs for the test set.

\subsubsection{Evaluation Metrics}
We estimate the quality of a static call graph using standard evaluation metrics: Precision, Recall, and F-measure, which are defined as follows:
\begin{equation}
 Precision = \dfrac{|S \cap G|}{|S|}
\end{equation}
\begin{equation}
 Recall = \dfrac{|S \cap G|}{|G|}
\end{equation}
\begin{equation}
 F-measure = \frac{2 \times Precision \times Recall}{Precision + Recall}
\end{equation}
where $S$ and $G$ are the edge set in the static call graph and ground-truth respectively.

Among these evaluation metrics, Precision is the proportion of edges in call graph that are true calls. 
A high Precision is desirable for reducing developer effort in inspecting false positives \cite{christakis2016developers}.
Recall refers to the proportion of true edges that are retained in call graph. 
Finally, we  consider F-measure, which is the harmonic mean of Precision and Recall. 
We use F-measure as a summary statistic to capture the tradeoff between Precision and Recall.

Following previous work \cite{utture2022striking}, we compute the average Precision, Recall, and F-measure for the evaluation set by taking the mean over Precision, Recall, and F-measure of individual programs. 
We also report the standard deviation of each metric.



\subsubsection{Client analyses}

A better static call graph should lead to practical improvements on client analyses using the call graph.
To assess the effect of the improvements to the call graph from \tool{}, we run experiments on two client analyses, null pointer analysis, and monomorphic call-site detection, using the call graph produced by WALA.
To perform a direct comparison, the client analyses selected are the same as those considered by Utture et al.~\cite{utture2022striking}.
For each client analysis, we compare \tool{} against the baseline that produced the best call graph.

\textbf{Null pointer analysis.} In the first client analysis, we use analysis by Hubert et al.~\cite{hubert2008semantic}, which is implemented in Wala \cite{fink2012wala}, to detect possible null pointer dereferences related to uninitialized instance fields based on the input static call graph.
This analysis is context-insensitive and field-insensitive.
Improving the analysis would reduce the amount of developer effort spent inspecting false alarms, which is known to be a barrier to the adoption of bug detection tools based on static analysis~\cite{johnson2013don}.
For this client analysis, we refer to incorrect warnings reported as ``false alarms'' to distinguish them from ``false positives'' from the call graph construction.
This analysis is independently checked by two human annotators to manually determine if the warnings are false alarms.
In cases where the two annotators
disagree on the decision, we involve a third annotator, an author of the paper, for a discussion to reach a consensus.
We report the total number of reported warnings and the false alarm rate of the null pointer analyzer.
As the ground-truth labels of all warnings are not known, we are not able to compute Recall and, therefore, we do not compare the approaches on F-measure.

\textbf{Monomorphic call-site detection.} In the second client analysis, the static call graph is used to detect monomorphic call sites. 
A call site is monomorphic when only one concrete function can be invoked.  
The detection of monomorphic call sites is useful, for example, for function inlining to reduce the runtime cost of function dispatch \cite{utture2022striking}.
A better call graph would allow us to safely inline more functions, improving the performance of programs.
The ground-truth of this analysis is determined by running the analysis on the ground-truth call graph.
We report the precision, recall, and F-measure of the monomorphic call site detector.


\begin{table}[htb]
        \caption{The classification threshold values of \cgpruner{} for different static call graphs}
        \centering
        \begin{tabular}{|l|c|c|c|}
             \hline
             \textbf{Call Graph} & \textbf{WALA} & \textbf{Doop} & \textbf{Petablox}  \\              \hline
             \textit{Balanced point} & 0.4500 & 0.4028 & 0.4279 \\
             \textit{Default} & 0.4500 & 0.5000 & 0.5000 \\              \hline
        \end{tabular}
        \label{tab:threshold_cgpruner}
\end{table}
    
\subsubsection{Baselines} \label{sec:exp_baseline}
To assess the effectiveness of \tool{}, we compare our approach with the following  baselines: 
\begin{itemize}
    \item the \textit{original} call graphs are call graphs constructed by static analysis tools. 
    In this work, we consider the 0-CFA static call graphs constructed by three standard  static analysis tools,  WALA \cite{fink2012wala}, Doop \cite{bravenboer2009strictly}, and Petablox \cite{mangal2015user}. The choice of these three analysis tools follows the previous work\cite{utture2022striking} for a fair comparison.
    \item a \textit{random} baseline that randomly removes $N$\% of edges in a call graph. For a fair comparison, we set $N$ as the percentage of edges that are removed by \tool{}.
    \item\cgpruner{}~\cite{utture2022striking} is the state-of-the-art technique in call graph pruning task. \cgpruner{} constructs a decision tree based on the 11 types of structural features (listed in Table \ref{tab:struc_feats}) extracted from the call graphs to identify false positive edges. 
    We run \cgpruner{} using the paper's replication package\footnote{https://doi.org/10.5281/zenodo.5177161}.
    \cgpruner{} uses a classification threshold to determine if an edge is a false positive. 
    A higher threshold results in higher precision as it accepts only a few edges which are more likely to be the ground-truth call graph. The threshold enables a trade-off between precision and recall.
    We report the result of \cgpruner{} at two different thresholds. 
    The first is determined based on the ``balanced point'', where the threshold is tuned such that the average precision and recall of the call graph are equal when evaluated on the test dataset, following the procedure used by Utture et al.~\cite{utture2022striking} to determine the optimal balance between precision and recall. 
    Note that, in the \cgpruner{} paper, the "balanced points" are identified using the evaluation dataset. In this paper, we selected the points using the training dataset to reduce likelihood of overfitting.
    \\
    The second threshold is the default threshold obtained from the replication package. 
    Note that, for WALA, both thresholds share the same value of 0.45, so we report just one set of results. The thresholds are shown in Table \ref{tab:threshold_cgpruner}.

    \item \textsc{GCN}~\cite{kipf2016semi} is an standard, off-the-shelf, graph neural network (GNN).
    In this task, we use it perform edge classification. GNNs are common for machine learning on graph-structured data. 
    As input, we pass the call graph constructed from static analysis along with 11 types of structural features described in Table \ref{tab:struc_feats}. 
    Then, we use the GCN, which considers information based on the nodes in the neighborhood, to predict if an edge in the call graph is a false positive.
    
\end{itemize}

\subsubsection{Implementation details}
For \tool{}, we implement the proposed approach using PyTorch library and  the Python programming language. 
The models are trained and evaluated on two NVIDIA RTX 2080 Ti GPU with 11GB of graphics memory.
For CodeBERT, we fine-tune the model with a learning rate of 1e-5, following prior works \cite{zhou2021finding, feng2020codebert} and a batch size of 10 in 2 epochs.
We trained the neural classifier with a learning rate of 5e-6 with a batch size of 50 in 5 epochs. 

\subsection{Experimental Results}

\subsubsection{Effectiveness of \textit{AutoPruner}}
\label{sec:rq1}

We report the comparison of our approach, \tool{} against the baselines approaches. 
The detailed results are shown in Table \ref{tab:overall}.

\begin{table}[htb]
\caption{Comparison of the effectiveness of \tool{} with the baselines on static call graph generated by WALA, Doop, and Petablox. $\text{\cgpruner{}}_{bal}$ and $\text{\cgpruner{}}_{def}$ denotes the result of \cgpruner{} at balanced point (where the precision and recall are equal on the test dataset) and default threshold (as provided in replication package), respectively. For WALA, these thresholds are the same, so we report only one set of results as \cgpruner{}. The bold and underlined number denotes the best result for F-measure.} \label{tab:overall}

\resizebox{\columnwidth}{!}{
\begin{tabular}{|l|l|c|c|c|}
\hline
\textbf{Tool} & \textbf{Technique} & \textbf{Precision} & \textbf{Recall}  & \textbf{F-measure}\\
\hline
& $original$  & 0.24 $\pm$ 0.21 & 0.95 $\pm$ 0.14 & 0.34 $\pm$ 0.24 \\
&\textit{random}& {0.24} $\pm$ 0.21 & 0.48 $\pm$ 0.07 & {0.27} $\pm$ 0.17 \\
WALA & \cgpruner{}  & 0.66 $\pm$ 0.19 & 0.66 $\pm$ 0.32 & 0.60 $\pm$ 0.25 \\
& \textsc{GCN} & 0.48 $\pm$ 0.2 & 0.74 $\pm$ 0.37 & 0.54 $\pm$ 0.29\\
& \tool{} & 0.69 $\pm$ 0.21 & 0.71 $\pm$ 0.19 & \underline{\textbf{0.68}} $\pm$ 0.19 \\ \hline
& $original$  & 0.23 $\pm$ 0.21 & 0.92 $\pm$ 0.14 &  0.33 $\pm$ 0.25 \\
&\textit{random}& {0.23} $\pm$ 0.22 & 0.46 $\pm$ 0.07 & {0.26} $\pm$ 0.17 \\
& $\text{\cgpruner{}}_{bal}$ &  0.67 $\pm$ 0.19 &  0.67 $\pm$ 0.30 & 0.61 $\pm$ 0.22 \\
Doop & $\text{\cgpruner{}}_{def}$ &  0.72 $\pm$ 0.19 &  0.53 $\pm$ 0.32 & 0.54 $\pm$ 0.26 \\
& \textsc{GCN} & 0.49 $\pm$ 0.23 & 0.77 $\pm$ 0.31 & 0.55 $\pm$ 0.25\\
& \tool{} & 0.64 $\pm$ 0.24 & 0.75 $\pm$ 0.15 & \underline{\textbf{0.66}} $\pm$ 0.19 \\ \hline
& $original$ & 0.30 $\pm$ 0.25 & 0.89 $\pm$ 0.14 &  0.40 $\pm$ 0.27 \\
&\textit{random}& {0.3} $\pm$ 0.24 & 0.45 $\pm$ 0.08 & {0.31} $\pm$ 0.18 \\
& $\text{\cgpruner{}}_{bal}$  &  0.67 $\pm$ 0.19 &  0.67 $\pm$ 0.27 & 0.61 $\pm$ 0.20 \\
Petablox & $\text{\cgpruner{}}_{def}$ &  0.73 $\pm$ 0.19 &  0.52 $\pm$ 0.35 & 0.52 $\pm$ 0.29 \\
& \textsc{GCN} & 0.52 $\pm$ 0.23 & 0.78 $\pm$ 0.28 & 0.58 $\pm$ 0.23\\
& \tool{} & 0.67 $\pm$ 0.21 & 0.69 $\pm$ 0.21 & \underline{\textbf{0.65}} $\pm$ 0.18 \\ \hline
\end{tabular}
}
\end{table}

The evaluation results demonstrate that \tool{} successfully boosts the performance of the call graphs produced by WALA, Doop, and Petablox by 
0.25--0.34, up to 100\% improvement ((0.68-0.34)/0.34) in F-measure.
Overall, the use of \tool{} led to gains in Precision (up to 178\% improvements) which are substantially larger than the slight losses in Recall (up to just 24\%).

With respect to the state-of-the-art baseline, \cgpruner{}, \tool{} further improves the baseline by 13\% in terms of F-measure for the call graph produced by WALA.
For the call graph of Doop and Petablox, our approach improves over the optimally balanced \cgpruner{} by 8\% and 7\%, respectively.
The improvements of \tool{} over \cgpruner{}  in F-measure are statistically significant (p-value < 0.05) using a Wilcoxon signed-rank test.

Note that the results above, obtained from $\text{\cgpruner{}}_{bal}$, is from \cgpruner{} with a classification threshold carefully tuned on the testing dataset to produce the optimal balance between precision and recall.
Therefore, $\text{\cgpruner{}}_{bal}$  represents the optimal performance of \cgpruner{} given the testing dataset.
It may not always be possible to obtain the 
optimal threshold in practice.
Despite that, we observe that \tool{} still outperforms \cgpruner{} on call graphs produced by all three static analysis tools with improvements in F-measure ranging from 7\%-13\%. 

When compared against $\text{\cgpruner{}}_{def}$, which is not optimally balanced on the testing dataset and uses the threshold listed in Table \ref{tab:threshold_cgpruner}, 
the improvements of \tool{} are more evident.
In terms of F-measure, \tool{} outperforms \cgpruner{} by 13\%--25\% .

Our approach performs better than the \textsc{GCN} baseline by
26\%, 20\%, and 12\% in terms of F-measure for call graph of WALA, Doop, and Petablox, respectively. 
Interestingly, \textsc{GCN} undeperforms \cgpruner{}. 
Both \textsc{GCN} and \cgpruner{} use only structural features from the call graph and differ only in the classifiers used (decision tree versus a graph neural network).
This shows that for our task of call graph pruning, the more complex classifier does not outperform the simpler classifier. 
One possible reason for this result is that the increased complexity of the \textsc{GCN} causes it to overfit the training data.
Overall, \tool{} outperforming both \textsc{GCN} and \cgpruner{} suggests that the  semantic features used by \tool{} have predictive power.




\begin{tcolorbox}
\textbf{\underline{Answer to RQ1:}}  Overall, \tool{}  outperforms every baseline approach, including the state-of-the-art call graph pruner. 
The call graphs pruned by \tool{} improves over the state-of-the-art baseline by up to 13\% in F1 when the baseline is optimally balanced and by up to 25\% when it is not.
Overall, \tool{} outperforms all baselines.
\end{tcolorbox}


\subsubsection{Effect on Client Analyses}
To investigate the effect of our pruned call graph on the client analyses (i.e., static analysis tools), we apply our call graph to two client analyses, i.e., null-pointer analysis and monomorphic call-site detection, following the experimental setup of prior work~\cite{utture2022striking}.

\begin{table}[htb]  
\caption{Comparison of the effectiveness of  \tool{} with the baselines and original call graph on null-pointer analysis. The bold and underline number denotes the best result for F-measure.} \label{tab:npe}
\begin{tabular}{|l|c|c|}
\hline
 \textbf{Techniques} &  \textbf{Total warnings} &  \textbf{False Alarms Rate} \\\hline
\textit{original}& {8,842} & {73\%}\\
\cgpruner{} & {757}  & {23\%}\\\hline
 \tool{}& {915}  & \underline{\textbf{12\%}} \\\hline
\end{tabular}
\end{table}

\vspace{0mm}\noindent \textbf{Null-pointer analysis.}
To investigate the impact of \tool{} in null pointer analysis, we pass the pruned call graph as input to a null pointer analysis~\cite{hubert2008semantic}, implemented in WALA. 
This analysis produces a set of warnings. 
Each warning is associated with a code location.
If the call graph is less accurate, the null-pointer analysis  produces  more false alarms.

To evaluate \tool{} and the baseline tools,
we use the same evaluation procedure as Utture et al.~\cite{utture2022striking} by performing manual analysis on the reported warnings. 
Specifically, one author (with four years of coding experience) and one non-author (with two years of coding experience) annotator independently manually inspect warnings produced by an analysis \cite{hubert2008semantic} implemented in WALA. 
A warning is considered a ``true alarm'' if the author can trace the backward slice of a dereference to an instance field that was uninitialized by the end of a constructor \cite{utture2022striking}. 
If another exception is encountered before dereferencing the null pointer, or if the label of a warning cannot be verified in 10 minutes or is otherwise unverifiable by the authors, then the warning is considered a ``false alarm''.
This labelling criteria for the human annotators  considers only the warnings produced by the program analysis~\cite{hubert2008semantic}, and therefore, is not a complete definition of a null pointer dereference. 
It is designed for a  analysis that is within the  cognitive ability of a human annotator to assess the call graphs produced by call graph pruners.
In the cases where the two annotators disagree on the decision, we involve a third annotator, an author of the paper (with three years of coding experience), for a discussion to reach a consensus.
Finally, we compute the false alarm rate of null-pointer analysis by dividing the number of false alarms by the number of warnings. The results are presented in Table \ref{tab:npe}. 
Furthermore, to assess the inter-rater reliability of the two annotators, we compute Cohen's Kappa~\cite{cohen1960coefficient} and obtained a value of 0.93, which is considered as almost perfect agreement~\cite{landis1977measurement}.

Using the call graph pruned by \tool{}, the null pointer analysis produces warnings with a false alarm rate of just 12\%.
Christakis and Bird~\cite{christakis2016developers} suggest that program analysis should aim for a false alarm rate no higher than 15-20\%, which is satisfied by \tool{}.
Meanwhile, both the original call graph and call graph produced by \cgpruner{} resulted in false alarm rates higher than 20\% (72\% and 23\% respectively). 
Our approach has a false alarm rate that is less than half of the \cgpruner{}'s false positive rate while reporting 158 more warnings.
With respect to the original call graph, our approach reduces the proportion of false alarms by six times, from 73\% to 12\%.

\vspace{0mm}

\begin{table}[htb]
\caption{Comparison of the effectiveness of  \tool{} with the baselines on monomorphic call-site detection using the call graph produced by WALA, Doop and Petablox. $\text{\cgpruner{}}_{bal}$ and $\text{\cgpruner{}}_{def}$ denotes the result of \cgpruner{} at balanced point (where the precision and recall are equal) and default threshold (as provided in the replication package), respectively. For WALA, these thresholds are the same, so we report only one set of results for  \cgpruner{}. The bold and underlined number denotes the best result for F-measure.} \label{tab:monomorphic}
\resizebox{\columnwidth}{!}{
\begin{tabular}{|l|l|c|c|c|}
\hline
\textbf{Tool}&\textbf{Techniques} & \textbf{Precision} & \textbf{Recall} & \textbf{F-measure} \\ \hline
&\textit{original}& 0.52 $\pm$ 0.23 & 0.93 $\pm$ 0.15 & {0.64} $\pm$ 0.21 \\
WALA &\cgpruner{}  & {0.68} $\pm$ 0.18 & {0.68} $\pm$ 0.32 & {0.62} $\pm$ 0.25 \\
&\tool{}& 0.71 $\pm$ 0.22 & 0.72 $\pm$ 0.19 & \underline{\textbf{0.69}} $\pm$ 0.19
\\ \hline
&\textit{original}& {0.51} $\pm$ 0.24 & 0.92 $\pm$ 0.14 & {0.63} $\pm$ 0.22 \\
Doop &$\text{\cgpruner{}}_{bal}$ & 0.68 $\pm$ 0.19 & {0.71} $\pm$ 0.31 & {0.63} $\pm$ 0.23 \\
&$\text{\cgpruner{}}_{def}$ & 0.68 $\pm$ 0.21 & {0.56} $\pm$ 0.37 & {0.53} $\pm$ 0.30 \\
&\tool{}& 0.66 $\pm$ 0.24 & 0.78 $\pm$ 0.16 & \underline{\textbf{0.69}} $\pm$ 0.20 \\ \hline
&\textit{original}& {0.52} $\pm$ 0.23 & 0.9 $\pm$ 0.15 & {0.63} $\pm$ 0.21\\
Petablox &$\text{\cgpruner{}}_{bal}$ & {0.68} $\pm$ 0.19 & {0.72} $\pm$ 0.28 & {0.63} $\pm$ 0.21 \\
&$\text{\cgpruner{}}_{def}$ & 0.73 $\pm$ 0.19 & {0.58} $\pm$ 0.34 & {0.56} $\pm$ 0.26 \\
&\tool{} & 0.69 $\pm$ 0.21 & 0.75 $\pm$ 0.20 & \underline{\textbf{0.68}} $\pm$ 0.19
\\ \hline
\end{tabular}
}
\end{table}

\noindent \textbf{Monomorphic call-site detection.} 
In the task of monomorphic call site detection, the call graph is used to identify call sites where there is only one concrete call at a code location.
As shown in Table \ref{tab:monomorphic}, for the call graph constructed by WALA,
\tool{} outperforms the original call graph and \cgpruner{} in F-measure.
\tool{} outperforms \cgpruner{} by 11\% in  F-measure, with improvements in both precision and recall.

We observe similar performances on the call graph of Doop and Petablox.
On Doop's call graph,
\tool{} outperforms \cgpruner{} by 30\% in F-measure. 
Compared to the original graph,  
\tool{} improves in F-measure by 10\%.
Similarly, on the call graph produced by Petablox, \tool{} improves over both \cgpruner{} and the original call graph by 8\% in terms of F-measure. 


\begin{tcolorbox}
\textbf{\underline{Answer to RQ2:}} The call graph produced by \tool{} leads to improvements in both null pointer analysis and monomorphic call site detection. 
Based on the call graph from WALA, \tool{} decreases the false alarm rate from null pointer analysis by 11\%.
On monomorphic call site detection, \tool{} improves over the state-of-the-art call graph pruner by 11\% in F-measure.
\end{tcolorbox}

\begin{figure*}
\centering
\begin{subfigure}[H]{0.33\textwidth}
\includegraphics[width=\textwidth]{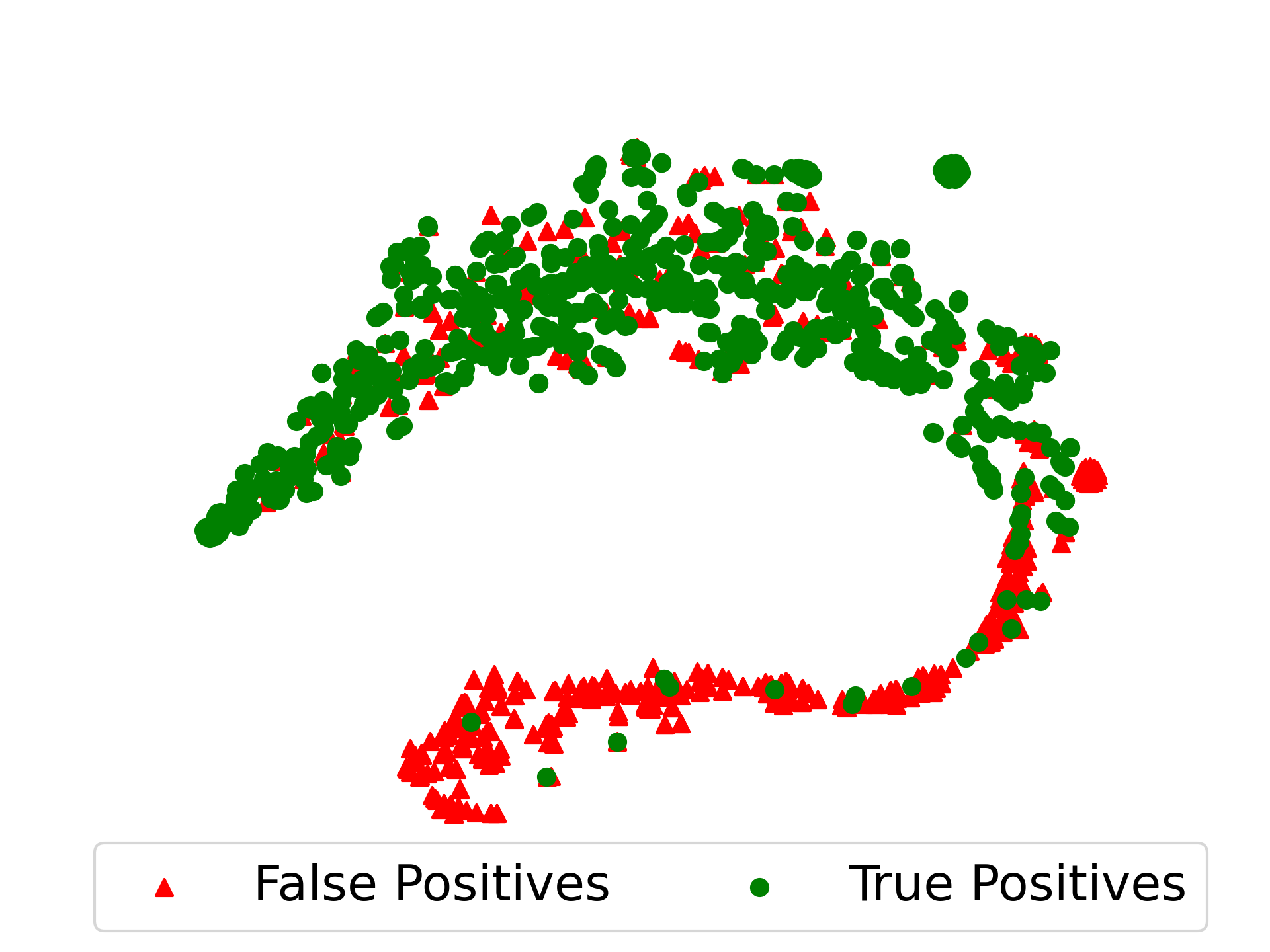}
\end{subfigure}
\begin{subfigure}[H]{0.33\textwidth}
\includegraphics[width=\textwidth]{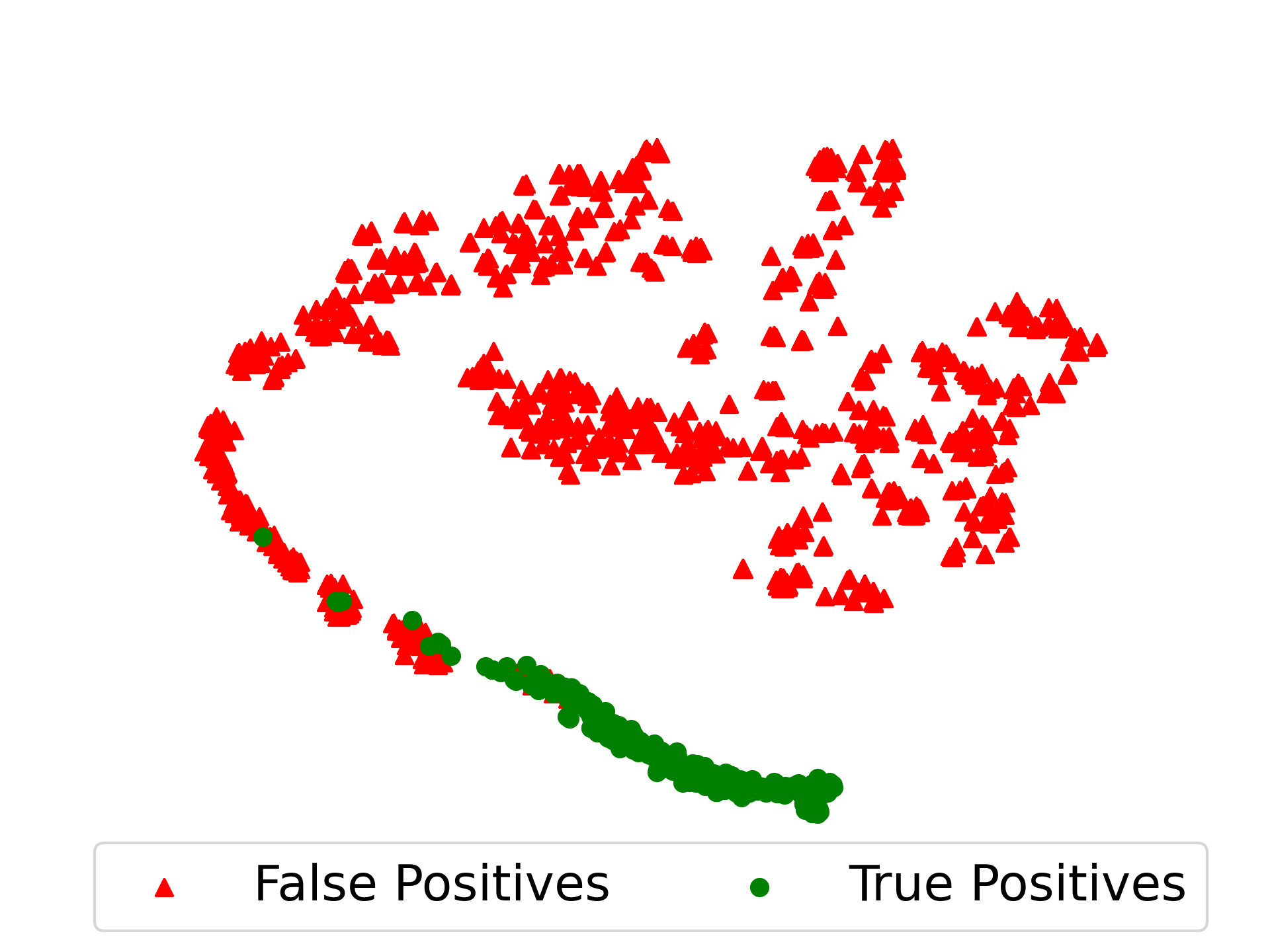}
\end{subfigure}
\begin{subfigure}[H]{0.33\textwidth}
\includegraphics[width=\textwidth]{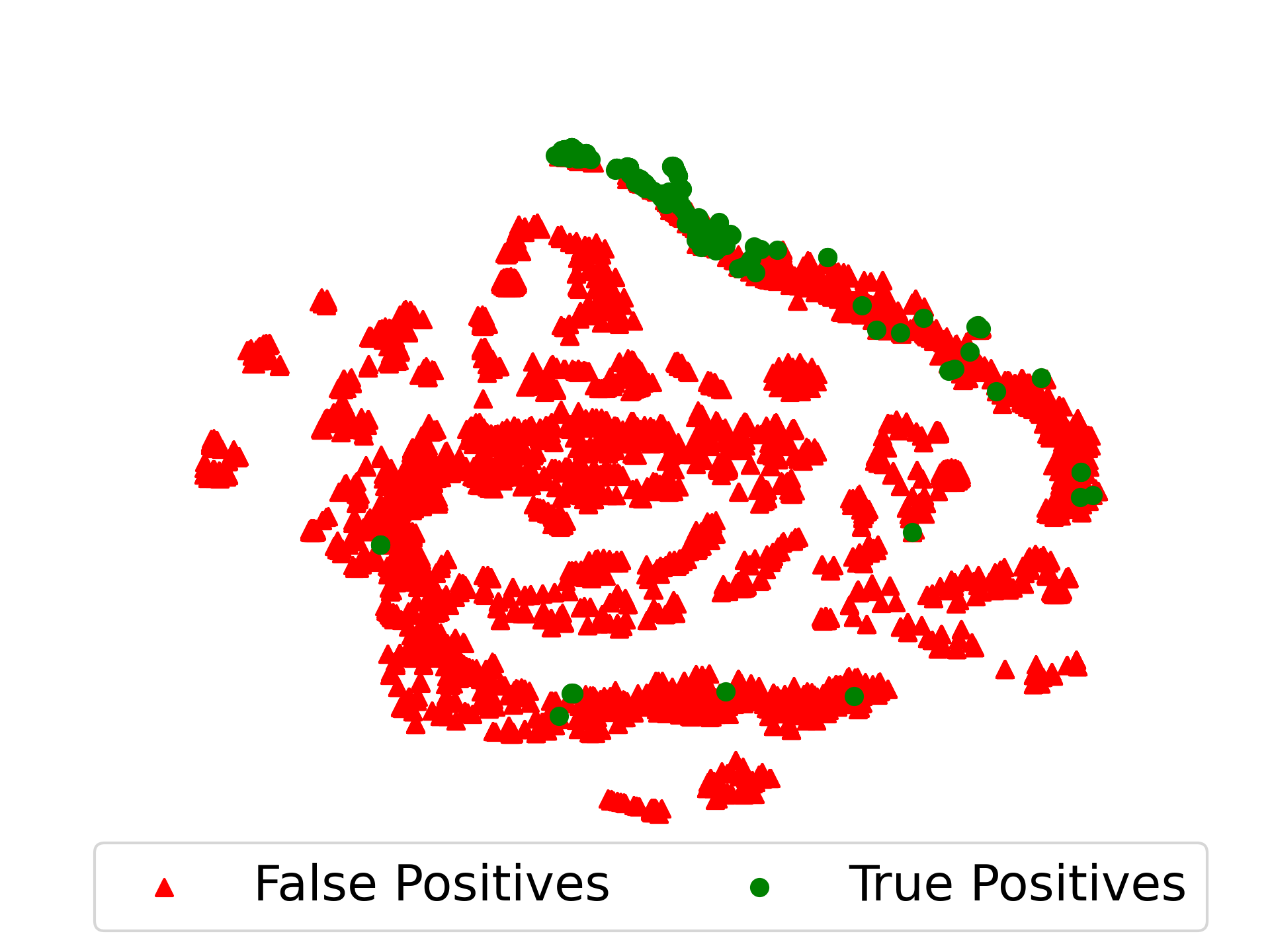}
\end{subfigure}
\begin{subfigure}[H]{0.33\textwidth}
\includegraphics[width=\textwidth]{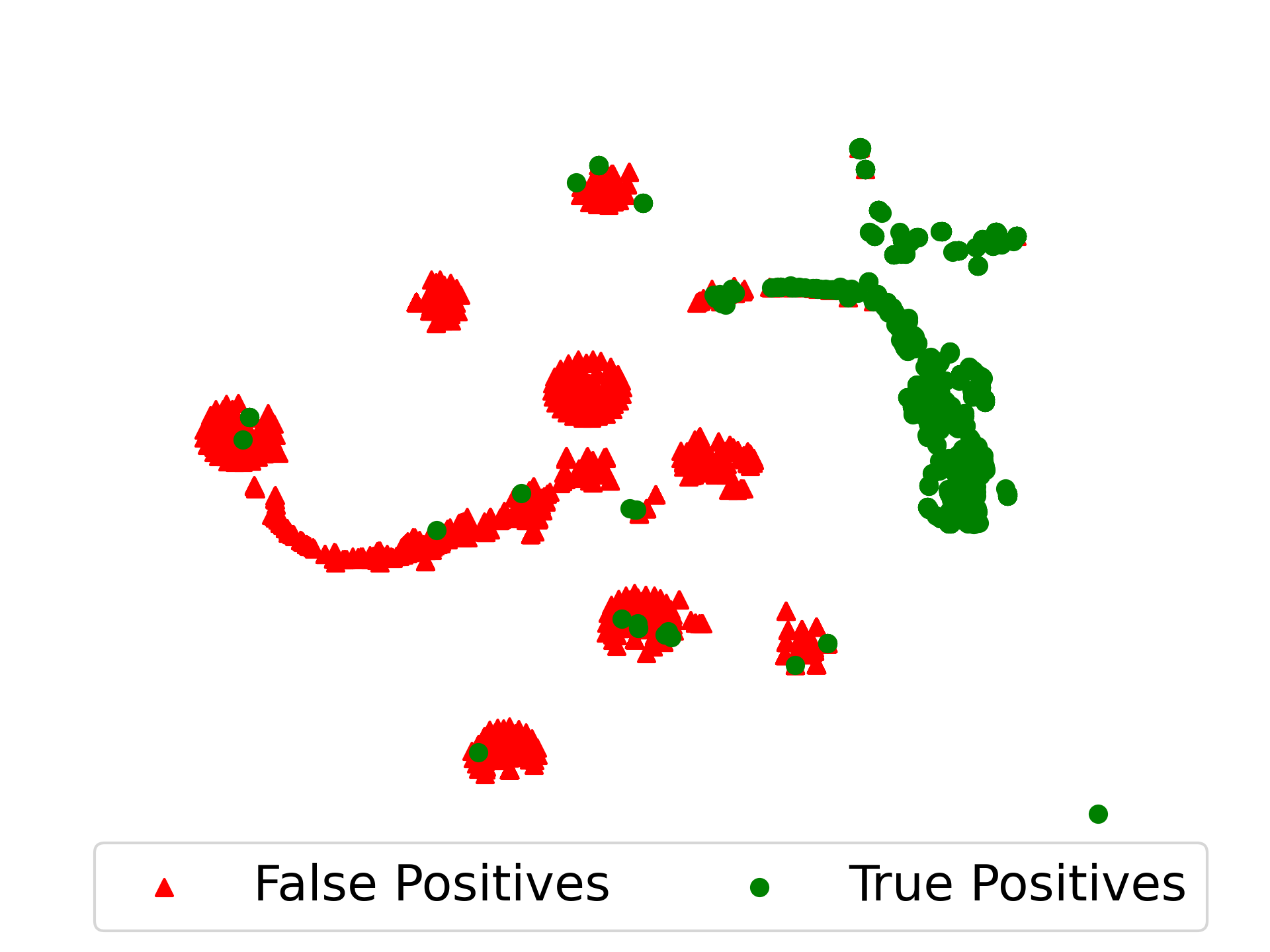}
\end{subfigure}
\begin{subfigure}[H]{0.33\textwidth}
\includegraphics[width=\textwidth]{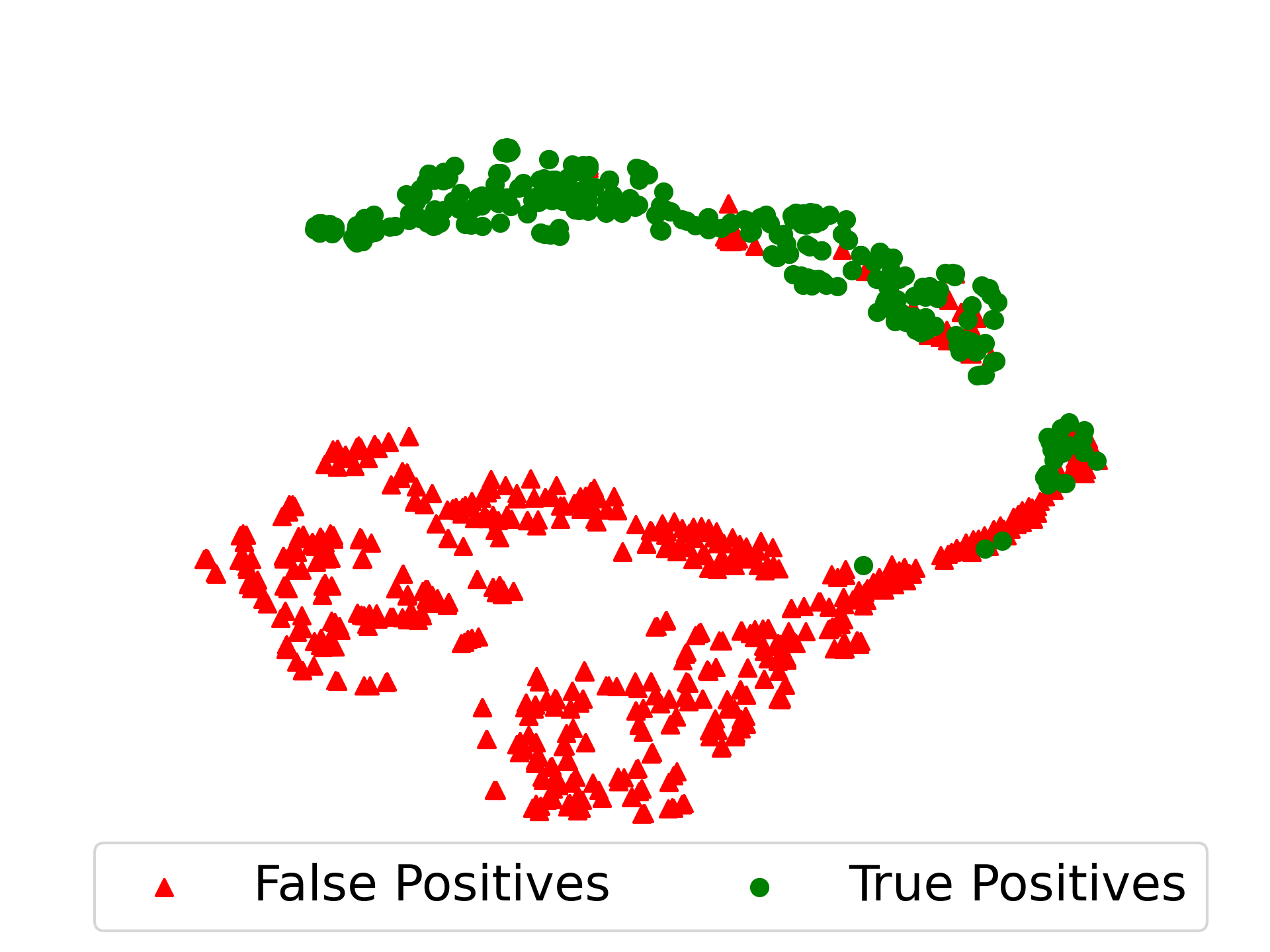}
\end{subfigure}
\begin{subfigure}[H]{0.33\textwidth}
\includegraphics[width=\textwidth]{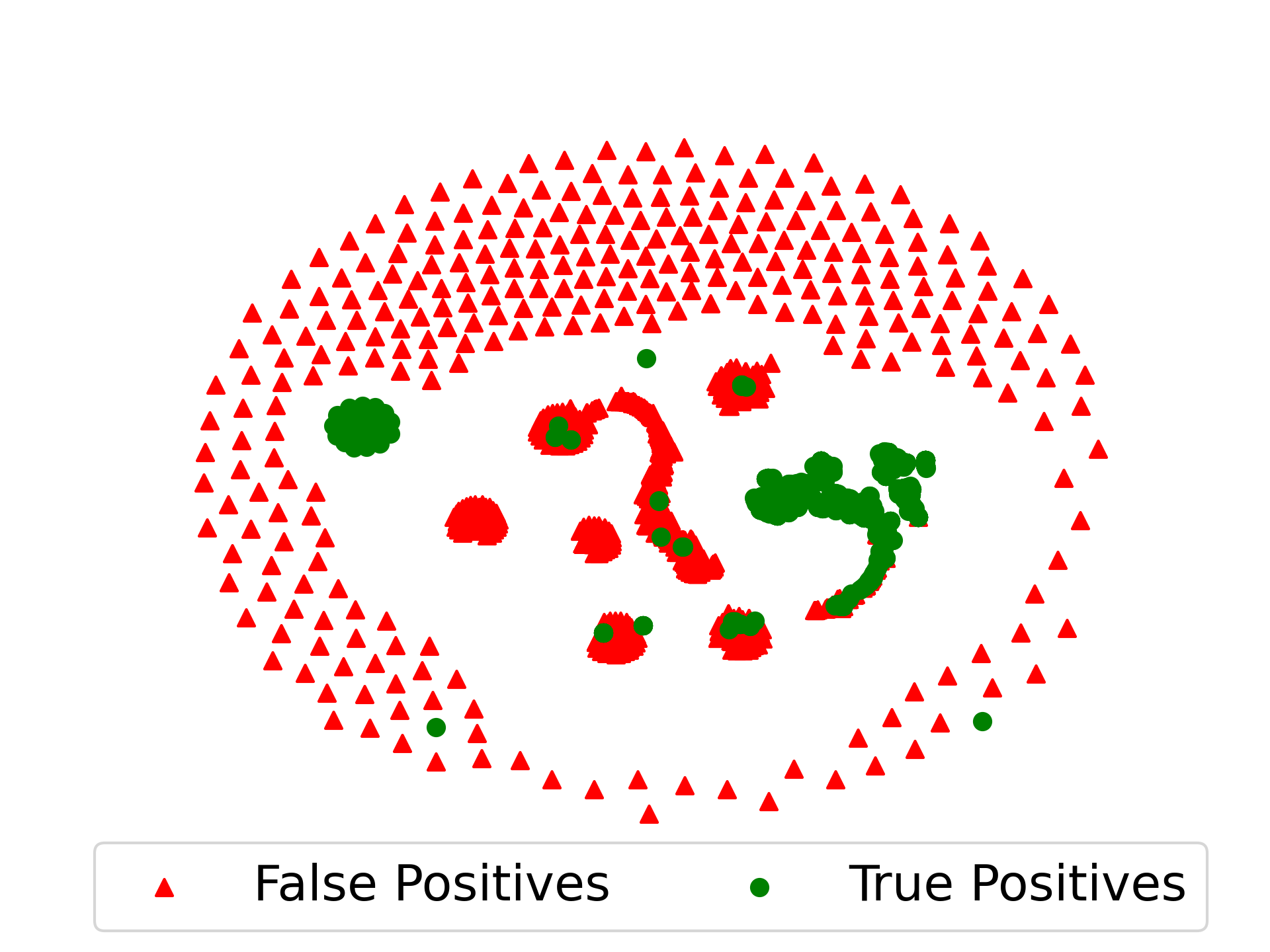}
\end{subfigure}
\caption{The visualization of semantic features produced by \textit{CodeBERT} in 6 sample programs. The \textcolor{officegreen}{green} circles and \textcolor{red}{red} triangles  represent true positive and false positive edges respectively.}
\label{fig:visual}
\end{figure*}

\subsubsection{Ablation Study} \label{sec:ablation}

In answer this question, we investigate two different ablation studies: 
\begin{itemize}
    \item Semantic versus Structure
    \item Caller versus Callee function
\end{itemize}

\begin{table}[htb]
\caption{Comparison of the effectiveness of  the semantic features and the structural features. The bold and underline number denotes the best result for F-measure.} \label{tab:ablation1}
\begin{tabular}{|l|c|c|c|}
\hline
\textbf{Techniques} & \textbf{Precision} & \textbf{Recall}  & \textbf{F-measure}\\
\hline
$\text{\tool{}}_{struct}$& 0.58 & 0.75 & 0.62 \\

$\text{\tool{}}_{sem}$& 0.67 & 0.71 & 0.66\\
\hline
 \tool{} & 0.69 & 0.71& \underline{\textbf{0.68}} \\ \hline
\end{tabular}
\end{table}

\vspace{0mm}

\noindent \textbf{Semantic vs. Structure.} In this experiment, we evaluate the relative contribution of  \tool{}'s semantic versus structural features for call graph pruning.
Table \ref{tab:ablation1} shows the results of our experiments.
$\text{\tool{}}_{sem}$ refers to \tool{} using only the semantic features extracted from the source code, and $\text{\tool{}}_{struct}$, refers to the \tool{} using only the structural features. 
Using only the semantic features, $\text{\tool{}}_{sem}$ outperforms \\ $\text{\tool{}}_{struct}$ in F-measure by 6\%. By using with both types of features, \tool{} outperforms $\text{\tool{}}_{sem}$ by 3\% in F-measure.
The decreases in F-measure when either the semantic features or structural features are removed are statistically significant (p-value < 0.05).
This indicates that both types of features are important, but relatively larger decrease in F-measure when removing the semantic features indicates that the semantic features are more important compared to the structural features.

Overall, when \tool{} uses only one type of feature,  \tool{} has a lower F-measure. 
This suggests that both semantic and structural features are important for \tool{} to perform effectively. 


\begin{table}[htb]
\caption{Comparison of the effectiveness of the caller features and the callee features. The bold and underline number denotes the best result for F-measure.} \label{tab:ablation2}
\begin{tabular}{|l|c|c|c|}
\hline
\textbf{Techniques} & \textbf{Precision} & \textbf{Recall}  & \textbf{F-measure}\\
\hline
$\text{\tool{}}_{caller}$& 0.69 & 0.60 & 0.58 \\

$\text{\tool{}}_{callee}$& 0.68 & 0.66 & 0.60\\
\hline
 \tool{} & 0.69 & 0.71 & \underline{\textbf{0.68}} \\ \hline
\end{tabular}
\end{table}

\vspace{0mm}

\noindent \textbf{Caller vs. Callee.} 
Next, we assess the relative importance of the features extracted from the caller and callee functions. 
We evaluate the performances of \tool{} when only considering source code from either caller and callee and compare them with \tool{}'s.
To perform this study, we fine-tuned CodeBERT with only either the source code of the caller or the callee function.

$\text{\tool{}}_{caller}$ refers to \tool{} using only the source code from the caller function, and $\text{\tool{}}_{callee}$ refers to \tool{} using only the source code from the callee function. 
\tool{} outperforms $\text{\tool{}}_{caller}$ and      $\text{\tool{}}_{callee}$ by up to 17\%.
The decreases in F-measure are statistically significant (p-value < 0.05).
This suggests that the semantic features extracted from the source code of both the caller and callee functions are crucial for the performance of \tool{}. 

\begin{tcolorbox}
\textbf{\underline{Answer to RQ3:}} Our ablation study shows that all features contribute to the effectiveness of \tool{}. The semantic features are more important than the structural features, while both the caller and callee functions are essential.   
\end{tcolorbox}

\section{Discussion}\label{sec:threats}
\subsection{Qualitative Analysis} \label{sec:qualitative}
In this section, we perform a qualitative analysis of \tool{}. 
We have seen that \tool{} consistently outperforms the state-of-the-art approach, and our ablation study indicates that the semantic features extracted by CodeBERT are essential for the effective performance of \tool{}.
Here, our goal is to investigate if the pre-trained transformer model of code, i.e., CodeBERT, is able to separate true positive edges from false positives in the call graph. 
To this end, we use \textit{t-SNE} \cite{van2008visualizing}, an unsupervised method for visualization, to visualize the semantic features from the call graphs of 6 programs in two-dimensional space. 
If the semantic features have predictive power, we would expect the call graph edges, which are false positives, to be separated from the true positives. 

Figure \ref{fig:visual} presents the visualizations, where the green and red points are features of true-positive and false-positive edges, respectively. 
Indeed, we observe that the majority of the green points are clustered and are located relatively close to one another. 
Furthermore, the clusters of green points are also typically far away from the majority of the red points. 
The observation suggests that \textit{CodeBERT} was able to be trained to extract semantic features from the caller and callee function such that the true positives can be separated from the false positives in the vector space.
This validates our use of a fine-tuned CodeBERT model for extracting semantic features, as the model demonstrates a remarkable ability to distinguish between true and false positives.

Still, a small but significant proportion of red and green points are located close to one another, indicating that using only semantic features is not enough to separate these edges. 
This suggests that other features (e.g., structural features) should be used to improve the model's classifier ability. 
Indeed, as shown in Section \ref{sec:ablation},
the structural features are complementary to the semantic features.
The addition of structural features increases the Precision of \tool{} by 3\% while keeping the same Recall, leading to 3\% improvement on F-Measure. 
Nevertheless, we acknowledge the modest contribution of the structural features which may be from how the structural and semantic features are combined. Currently, after they are independently extracted, they are combined with a small feed-forward neural network (FFNN). As a result, the FFNN may fail to capture more complex interactions between the semantic and structural features.

\subsection{Efficiency}

In this section, we investigate the efficiency of \tool{}. For pre-training model, \tool{} uses an existing pretrained model, CodeBERT\footnote{https://github.com/microsoft/CodeBERT}. For the offline fine-tuning (Section \ref{sec:finetune}) and training (Section \ref{sec:training}) phase wherein both the CodeBERT and the binary classifier needs to be finetuned and trained only once, \tool{} takes around 36 hours. For the inference phase which is integrated in downstream applications, \tool{} takes around 0.04 second on average to predict a label for an edge.

\subsection{Threats to validity}
\subsubsection{External validity} Threats to external validity concern the generalizability of our findings. 
Our experiments are performed on the same dataset as prior work~\cite{utture2022striking}, constructed from the NJR-1 benchmark\cite{palsberg2018njr}. 
This may be a threat to external validity since \tool{} may not generalize beyond the programs outside the NJR-1 dataset. 
However, this threat is minimal as the dataset consists of a large number of data points, and the NJR-1 benchmark is large and was carefully constructed to ensure their diversity~\cite{palsberg2018njr}.

\subsubsection{Internal validity} Threats to internal validity refer to possible errors in our experiments. 
In this study, following the experimental procedure of prior work \cite{utture2022striking}, we perform a manual inspection of null-pointer analysis, which may introduce human error. 
To minimize the risk, we asked one author of this paper and a non-author to independently inspect and label the reported warnings. 
We have measured the inter-rater reliability, obtaining Cohen's kappa of 0.93, which can be  interpreted as almost perfect agreement~\cite{landis1977measurement}. Therefore, we believe that there are minimal threats from this issue.


\subsubsection{Construct validity} Threats to construct validity relate to the suitability of our evaluation.
To minimize risks to construct validity, we have used the same experimental setup as a previous study~\cite{utture2022striking}, including the same dataset and ground-truth labels.
Some bias could be introduced in the construction of the ground truth. 
Manual labeling requires extensive human effort, which limits the scale of the experiments. 
Hence, we use the same  automated labeling procedure from prior work~\cite{utture2022striking} that runs test cases to identify ground truth edges. 
The imperfect code coverage of the test cases may introduce bias.
However, the code coverage in our experiments (68\%) is higher than the code coverage of real-world programs (less than 40\%) reported in prior studies~\cite{kochhar2017code}.

Another threat to construct validity is the definition of null pointer analysis of warnings produced by \cite{hubert2008semantic}. For analyzing static analysis warnings, prior works often employ a human study with several annotators \cite{tomassi2021real, habib2018many}. This human study is expensive and the task given to the annotators must be within the cognitive ability of humans. To make the annotation task tractable to humans and reduce its cost, we use the true-alarm identification task definition used in the prior work\cite{utture2022striking}.
\section{Related Work}\label{sec:related}
In Section \ref{sec:background}, we have discussed the studies related to call graph pruning and CodeBERT. 
Here, we discuss other related studies.

Call graph construction has been widely studied.
As our approach does not use runtime information, it falls into the class of static approaches~\cite{murphy1998empirical,utture2022striking,sui2020recall,reif2019judge} for constructing call graphs.
Approaches that use dynamic analysis~\cite{xie2002empirical, hejderup2018software} result in fewer false positives but are less scalable.

Some recent studies on call graph construction have focused on dynamic languages.
Salis et al.~\cite{salis2021pycg} and Nielson et al.~\cite{nielsen2021modular} present approaches for constructing call graphs of Python programs and Javascript programs.
Unlike language-specific techniques, AutoPruner can be used to improve call graphs of any language.

There are many client analyses using call graphs. 
Recently, call graphs have been used for practical applications, including scanning applications for vulnerable library usage~\cite{nielsen2021modular}, generating exploits of vulnerabilities~\cite{iannone2021toward}, and impact analysis~\cite{hejderup2018software}.
We have explored two classic client analyses, null pointer analysis and monomorphic call site detection, to validate that the improvements from pruning the call graph lead to further improvements in practical applications.

Apart from applying CodeBERT to call graph contruction, researchers have proposed other applications of deep learning models on source code.
Other researchers had success using deep learning models for type inference~\cite{hellendoorn2018deep,allamanis2020typilus,pradel2020typewriter,peng2022static}, 
code completion~\cite{hindle2016naturalness,raychev2014code,svyatkovskiy2019pythia}, code clone detection~\cite{saini2018oreo}, program repair~\cite{mashhadi2021applying, chen2019sequencer}, fault localization \cite{nguyen2022ffl, lou2021boosting}, among other analyses on source code~\cite{leclair2019neural,lu2021codexglue,bui2021infercode, nguyen2022vulcurator}.
Our work is similar as AutoPruner successfully applies deep learning techniques on source code analysis, 
but is unique as previous studies have not previously applied deep learning to call graph analysis.

\section{Conclusion and Future Work}\label{sec:conclusion}
We propose \tool{}, a novel call graph pruner that leverages both structural and semantic features. 
\tool{} employs CodeBERT to extract semantic features from both the caller and callee function associated with each edge in the call graph.
Our empirical evaluation shows that \tool{} outperforms multiple baselines, including the state-of-the-art approach that uses only structural features.
The improvements from \tool{} also lead to tangible improvements on downstream applications. In particular, the proportion of false alarms reported by a baseline null pointer analysis is halved, decreasing from 23\% to just 12\%.

Our ablation study shows that the semantic features complement the structural features. Moreover, our qualitative analysis reveals that the semantic features extracted by CodeBERT effectively separate true and false positive edges.

In the future, we will evaluate \tool{} on call graphs constructed using additional static analysis tools and for other programming languages, as well as assess the impact on call graph pruning on other client analyses. 
We will also explore more ways to improve \tool{}, such as jointly extracting semantic and structural features,
and providing more contextual information by incorporating the $k$-hop callers (e.g. the caller of the caller) of each function call to enrich the semantic features. 
Another interesting direction is to use \tool{} for proposing edges in the call graph missing due to an unsoundness program analysis. 


\vspace{2mm}

\noindent \textbf{Data availability.} AutoPruner’s dataset and implementation are publicly available at \url{https://github.com/soarsmu/AutoPruner/}.

\section*{Acknowledgement}

This project is supported by the National Research Foundation, Singapore and National University of Singapore through its National Satellite of Excellence in Trustworthy Software Systems (NSOE-TSS) office under the Trustworthy Computing for Secure Smart Nation Grant (TCSSNG) award no. NSOE-TSS2020-02. Any opinions, findings and conclusions or recommendations expressed in this material are those of the author(s) and do not reflect the views of National Research Foundation, Singapore and National University of Singapore (including its National Satellite of Excellence in Trustworthy Software Systems (NSOE-TSS) office). 

Xuan-Bach D. Le is supported by the Australian Government through the Australian Research Council’s Discovery Early Career Researcher Award, project number DE220101057.


\balance
\bibliographystyle{ACM-Reference-Format}
\bibliography{sample-base.bib}
\end{document}